\documentstyle[amssymb,aps,epsfig,twocolumn]{revtex}

\begin{document}

\title{Scalable photonic quantum computation through cavity-assisted interaction}
\author{L.-M. Duan$^{1}$ and H. J. Kimble$^{2}$}
\address{$^{1}$Department of Physics and FOCUS Center,
University of Michigan, Ann Arbor, MI 48109-1120\\
$^{2}$Norman Bridge Laboratory of Physics 12-33, California
Institute of Technology, Pasadena, CA 91125}
\maketitle

\begin{abstract}
We propose a scheme for scalable photonic quantum computation
based on cavity assisted interaction between single-photon pulses.
The prototypical quantum controlled phase-flip gate between the
single-photon pulses is achieved by successively reflecting them
from an optical cavity with a single-trapped atom. Our proposed
protocol is shown to be robust to practical nose and experimental
imperfections in current cavity-QED setups.
\end{abstract}


Realization of quantum computation requires accurate coherent control of a
set of qubits. A small volume optical cavity provides a platform to achieve
strong coherent interactions between atoms and photons, and has been
exploited as the critical component in several schemes for implementation of
quantum computation and communication \cite{pellizzari95,cirac97,monroe02}.
In a prototypical cavity-based quantum computation scheme of Ref. \cite
{pellizzari95}, the atoms are adopted as qubits while photons mediate the
interaction between them. Scaling to large-scale quantum computation via
this paradigm then requires that many atoms be localized and separately
addressed within a tiny optical cavity \cite{pellizzari95}, or alternatively
be coherently transported into and out of the cavity mode \cite{chapman}.
However, in spite of recent significant laboratory advances \cite
{mckeever03,keller03,chapman,rowe02,kuhn02}, these tasks remain daunting
experimental challenges.

Here, we propose a scalable quantum computation scheme where qubits are
encoded as polarizations of single-photon pulses. An optical cavity with a
single trapped atom is employed as the critical resource to achieve
controlled gate operations between photonic qubits and to act as a high
efficiency single-photon detector. The proposed computation architecture is
based on the state-of-the-art in cavity quantum electrodynamics \cite
{mckeever03}, can be readily scaled up to many qubits, and could be
integrated with protocols for the realization of quantum networks \cite
{cirac97}.

Quantum computation with single-photon polarizations as qubits \cite
{chuang95,turchette95} has the obvious advantage that the number of qubits
can readily be scaled up by generating many single-photon pulses. The main
obstacle to this approach is that it is exceedingly difficult to achieve
quantum gate operations between single-photon pulses. The typical
photon-photon coupling rate in available materials is orders of magnitude
too small to allow for any meaningful gate operation at the single-quantum
level. An interesting idea, as has been put forward recently in the
so-called linear optics quantum computation scheme \cite{knill01}, is to
achieve effective nonlinear interaction between photons through feed-forward
from high efficiency single-photon detectors. Though this approach is a very
important advance, a significant obstacle is that the required efficiency $%
\alpha $\ of the single-photon detectors for scalable quantum computing is
extremely high (e.g., for gate success with probability $p\simeq 0.99$, $%
\alpha \gtrsim 0.999987$ \cite{glancy02}).

In our proposed scheme, we combine the advantage of scalability from the
photonic qubits and the power of strong atom-photon coupling in a
high-finesse optical resonator. Such a cavity with one or few atoms in a
configuration of far-off-resonant interactions provides an effective Kerr
nonlinearity for the input light \cite{turchette95,imamoglu97,gheri98}, as
was first observed in Ref. \cite{turchette95}. However, this nonlinear phase
shift is typically too small for realization of the\ operation of the
prototypical quantum Controlled-NOT gate (C-NOT). Compared with the approach
of Ref. \cite{turchette95}, our new protocol has the following significant
advances: {\it (i)} A different interaction mechanism between photon pulses
leads to a much larger effective interaction rate sufficient for the
realization of a quantum C-NOT gate with current experimental capabilities.
{\it (ii)} The conditional phase flip in our scheme is very insensitive to
variation of the atom-photon coupling rate, so that high-fidelity gate
operations can be realized even if the atom is not localized in the
Lamb-Dicke regime. {\it (iii)} The pulse shapes for pairs of interacting
single photons suffer very small changes due to interactions with the
atom-cavity system, which is otherwise quite difficult to achieve \cite
{gheri98}. {\it (iv)} Finally, the noise properties of our scheme are quite
favorable, and should allow significant improvement in the error threshold
for large-scale, fault-tolerant quantum computation.

The basis states for our qubit consist of two orthogonal polarization states
of a single-photon pulse, denoted by $\left| h\right\rangle $ and $\left|
v\right\rangle $. A series of single-photon pulses is generated by emission
from a single atom in a cavity \cite{law97,cirac97}; single-qubit operations
on these photonic qubits are accurately performed through polarization
rotations. The critical problem for quantum computation with these qubits is
to achieve a nontrivial two-qubit interaction. Here, we choose the quantum
controlled phase flip (CPF), where the CPF\ gate for qubits $j$ and $k$
flips the phase of the input state if both qubits are in $\left|
h\right\rangle $ polarizations, and has no effect otherwise. The CPF gates,
together with simple single-qubit operations, realize universal quantum
computation \cite{preskill-notes}.

As illustrated in Fig. 1a, the CPF\ gate for two arbitrary pulses $j$ and $k$
is implemented by simply reflecting them successively from a high-Q cavity
which contains a single-trapped atom. The atom has three relevant levels as
shown in Fig. 1b, and is initially prepared in an equal superposition of the
two ground states, i.e., $\left| \Phi _{ai}\right\rangle =\left( \left|
0\right\rangle +\left| 1\right\rangle \right) /\sqrt{2}$. The atomic
transition $\left| 1\right\rangle \longrightarrow \left| e\right\rangle $ is
resonantly coupled to a cavity mode $a_{h}$, which has $h$ polarization and
is resonantly driven by the $h$ polarization component of the input
single-photon pulse. The $v$ polarization component of the input pulse is
reflected by the mirror M.

\begin{figure}[tb]
\epsfig{file=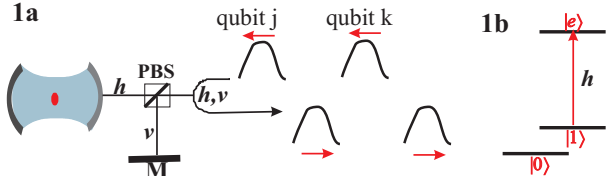,width=8cm}
\caption{(a) Schematic setup to implement the controlled phase
flip (CPF) gate between two single-photon pulses $j$ and $k$. With
a polarization beam splitter (PBS), the $h$-polarized component of
the single-photon pulse is reflected by the cavity, while the
$v$-polarized component is reflected via the mirror $M$. The
optical paths from the polarization beam splitter (PBS) to the
cavity and to the mirror $M$ are assumed to be equal. (b) The
relevant level structure of the atom trapped in the cavity (e.g.,
the states $\left\vert 0\right\rangle $ and $\left\vert
1\right\rangle $ could denote hyperfine states of an alkali atom
in the ground-state manifold while $\left\vert e\right\rangle $ is
an excited state).} \label{fig1}
\end{figure}

Before describing the detailed model and supporting calculations, first we
summarize the basic ideas of our scheme, which consists of two critical
steps. {\it (A)} By reflecting one single-photon pulse, say $j$, from the
cavity and the mirror, a CPF gate between the atom and the pulse $j$ is
achieved as described by the unitary operator $U_{aj}^{CPF}=e^{i\pi
\left\vert 0\right\rangle _{a}\left\langle 0\right\vert \otimes \left\vert
h\right\rangle _{j}\left\langle h\right\vert }$. {\it (B)} A composition of
the CPF gates between the atom and the pulses $j,k$ generates a CPF\ gate
between the pulses $j$ and $k$ described by the unitary operator $%
U_{jk}^{CPF}=e^{i\pi \left\vert h\right\rangle _{j}\left\langle h\right\vert
\otimes \left\vert h\right\rangle _{k}\left\langle h\right\vert }$, while
restoring the atom into its initial state $\left\vert \Phi
_{ai}\right\rangle $. Experimentally the composition is performed by
successively \textquotedblleft bouncing\textquotedblright\ the pulses from
the cavity (see Fig. 1a).

{\it Step (A) -- }When the incoming photon is $v$ polarized, it will be
reflected by the mirror {\it M} without any phase and shape change. When the
incoming photon is in $h$ polarization, it is resonant with the bare cavity
mode if the atom is in the $\left| 0\right\rangle $ state and thus acquires
a phase of $e^{i\pi }$ after its reflection; however, if the atom is in the $%
\left| 1\right\rangle $ state, the frequency of the dressed cavity mode from
the resonant atom-cavity coupling is significantly detuned from the
frequency of the incoming pulse. In this case, the cavity functions in the
same fashion as the mirror {\it M} and the photon pulse is reflected without
a phase change. A composition of the above sub-processes realizes the
desired CPF gate $U_{aj}^{CPF}$ between the atom and the photon.

{\it Step (B) -- }Critical to the second step of our protocol is the
following operator identity:
\begin{eqnarray}
U_{jk}^{CPF}\left| \Psi _{jk}\right\rangle \otimes \left| \Phi
_{ai}\right\rangle  &=&U_{aj}^{CPF}R_{a}\left( -\pi /2\right)
U_{ak}^{CPF}R_{a}\left( \pi /2\right)   \nonumber \\
&&\times U_{aj}^{CPF}\left| \Psi _{jk}\right\rangle \otimes \left| \Phi
_{ai}\right\rangle \text{ ,}
\end{eqnarray}
where $\left| \Psi _{jk}\right\rangle $ denotes an arbitrary state of the
photonic qubits $j$ and $k$, and $R_{a}\left( \theta \right) $ is a
single-bit rotation on the atom which transforms according to $R_{a}\left(
\theta \right) \left| 0\right\rangle =\cos \theta /2\left| 0\right\rangle
+\sin \theta /2\left| 1\right\rangle $ and $R_{a}\left( \theta \right)
\left| 1\right\rangle =-\sin \theta /2\left| 0\right\rangle +\cos \theta
/2\left| 1\right\rangle $. The identity (1) demonstrates that the CPF\ gate
between two arbitrary single-photon pulses $j$ and $k$ can be implemented by
first reflecting the pulse $j$ from the cavity as shown in Fig. 1a, then
applying a $\left( \pi /2\right) $-pulse laser on the atom, then reflecting
the pulse $k$ from the cavity, then applying a $\left( -\pi /2\right) $%
-pulse laser on the atom, and finally reflecting the pulse $j$ again from
the cavity.

The CPF\ gate $U_{aj}^{CPF}$ between the atom and the photon pulse can also
be used to achieve quantum non-demolition (QND) measurement of the photon
number in the pulse. For this purpose, we simply prepare the atom in the
state $\left| \Phi _{ai}\right\rangle $, reflect the to-be-measured photon
pulse from the cavity, apply a $R_{a}\left( \pi /2\right) $ rotation on the
atom, and finally perform a measurement of the atomic state in the basis $%
\left\{ \left| 0\right\rangle ,\left| 1\right\rangle \right\} $.
The measurement outcome is ``$0$''\ if and only if the $h$
component of the pulse has a photon. By the same avenue, we can
also measure the parity of several photonic qubits (``parity''\
concerns whether a series of pulses has a total even or odd photon
number in their $h$ components) by successively reflecting them
from the cavity, and can as well measure the total photon number
of both $h$ and $v$ components of a single pulse by reflecting it
twice from the cavity with a polarization flip between the two
reflections. Such QND\ measurements have wide applications for
quantum information processing \cite{b-k98,duan00}. Note that the
measurement of atomic internal states can be done with near
$100\%$ efficiency through the quantum jump technique
\cite{monroe02}. So, the efficiency of our QND measurement is
principally only limited by the inefficiency of the CPF\ gate
between the atom and the photon pulse caused by atomic spontaneous
emission loss, which as we will see later, is significantly less
than the inefficiency of conventional destructive single-photon
detectors.

Now we present a detailed theoretical model to demonstrate that the CPF\
gate $U_{aj}^{CPF}$ between the atom and the single-photon pulse $j$ can be
obtained simply by reflecting the latter from the cavity. The initial state
of the pulse $j$ can be expressed as $\left| \Psi _{p}\right\rangle
_{j}=c_{hj}\left| h\right\rangle _{j}+c_{vj}\left| v\right\rangle _{j}$,
where $c_{hj}$ and $c_{vj}$ are arbitrary superposition coefficients. The
polarization component states $\left| \mu \right\rangle _{j}$ $\left( \mu
=h,v\right) $ have the form $\left| \mu \right\rangle
_{j}=\int_{0}^{T}f_{j}\left( t\right) a_{\mu }^{in\dagger }\left( t\right)
dt\left| \text{vac}\right\rangle $, where $f_{j}\left( t\right) $ is the
normalized pulse shape as a function of time $t$, $T$ is the pulse duration,
$a_{\mu }^{in}\left( t\right) $ are one-dimensional field operators (cavity
input operators) with the standard commutation relations $\left[ a_{\mu
}^{in}\left( t\right) ,a_{\mu ^{\prime }}^{in\dagger }\left( t^{\prime
}\right) \right] =\delta _{\mu \mu ^{\prime }}\delta \left( t-t^{\prime
}\right) $ \cite{walls94}, and $\left| \text{vac}\right\rangle $ denotes the
vacuum of all the optical modes. The cavity mode $a_{h}$ is driven by the
corresponding cavity input operator $a_{h}^{in}\left( t\right) $ through
\cite{walls94}
\begin{equation}
\dot{a}_{h}=-i[a_{h},H]-\left( i\Delta +\kappa /2\right) a_{h}-\sqrt{\kappa }%
a_{h}^{in}\left( t\right) \text{ ,}
\end{equation}
where $\kappa $ is the cavity (energy) decay rate and the Hamiltonian
\begin{equation}
H=\hbar g\left( \left| e\right\rangle \left\langle 1\right| a_{h}+\left|
1\right\rangle \left\langle e\right| a_{h}^{\dagger }\right)
\end{equation}
describes the coherent interaction between the atom and the cavity mode $%
a_{h}$. The detuning $\Delta $ in Eq. (2) is meant to be $0$ for our scheme,
but we retain it here for subsequent pedagogical purposes. The cavity output
$a_{h}^{out}\left( t\right) $ is connected with the input by the standard
input-output relation
\begin{equation}
a_{h}^{out}\left( t\right) =a_{h}^{in}\left( t\right) +\sqrt{\kappa }a_{h}%
\text{ .}
\end{equation}
As the $v$ component of the pulse is reflected by the mirror $M$, we simply
have $a_{v}^{out}\left( t\right) =a_{v}^{in}\left( t\right) $.

Equations (2)-(4) determine the evolution of the joint state of atom and
photon pulse, and can be solved without further approximation through
numerical simulation. However, before presenting the simulation results,
first we attack this problem analytically with some rough approximations to
reveal the underlying physics. If the atom is in the state $\left|
0\right\rangle $, the Hamiltonian $H$ does not play a role in Eq. (2). In
this case, from Eqs. (2) and (4) we find
\begin{equation}
a_{h}^{out}\left( t\right) \approx \frac{i\Delta -\kappa /2}{i\Delta +\kappa
/2}a_{h}^{in}\left( t\right) \text{ ,}
\end{equation}
where the high-frequency components of the field operators $a_{\mu
}^{in}\left( t\right) $ and $a_{\mu }^{out}\left( t\right) $ have
been
discarded, which is a valid approximation if the input pulse shape $%
f_{j}\left( t\right) $ changes slowly with time $t$ compared with the cavity
decay rate, i.e., $\left| \partial _{t}f_{j}\left( t\right) /f_{j}\left(
t\right) \right| \ll \kappa $. Under this approximation, we have $%
a_{h}^{out}\left( t\right) \approx -a_{h}^{in}\left( t\right) $ for resonant
interaction $\Delta =0$, so the $h$ component acquires the phase $\pi $
after reflection from the cavity. However, if the atom is in the state $%
\left| 1\right\rangle $, the response function of the cavity is modified by
the coupling (3), where for the case of strong coupling \cite{hjk98}, the
two dressed cavity modes have frequencies that are effectively detuned from
that of the input pulse by $\Delta =\pm g$, respectively. In the case that $%
g\gg \kappa $, we have $a_{h}^{out}\left( t\right) \approx a_{h}^{in}\left(
t\right) $ from Eq. (5), thereby confirming the preceding analysis to give
the desired CPF\ gate $U_{aj}^{CPF}$.

Armed with this understanding, we finally present exact numerical
simulations for the theoretical model described by Eqs. (2)-(4).
In the simulation, we discretize the continuum field operators
$a_{h}^{in}\left( t\right) $ and $a_{h}^{out}\left( t\right) $,
and change the dynamics into the Schrodinger picture to avoid
operator ordering. The details of the simulation method can be
found in Ref. \cite{duan03}. Atomic spontaneous emission noise is
effectively described by an imaginary part $\left( -i\gamma
_{s}/2\right) \left( \left| e\right\rangle \left\langle e\right|
-\left| 1\right\rangle \left\langle 1\right| \right) $ in the
Hamiltonian $H$ \cite{duan03}, where $\gamma _{s}$ is the
spontaneous emission rate from the
state $|e\rangle $. The input pulse is taken to be Gaussian with $%
f_{j}\left( t\right) \propto \exp \left[ -\left( t-T/2\right) ^{2}/\left(
T/5\right) ^{2}\right] $ , where $t$ ranges from $0$ to $T$.

The numerical simulations show that the CPF\ gate $U_{aj}^{CPF}$ works
remarkably well. First of all, the conditional phase factor is either $%
e^{i\pi }$ or $e^{i0}$ depending on the atomic state $\left|
0\right\rangle $ or $\left| 1\right\rangle $, and this phase
factor is very insensitive to the variation of the coupling rate
$g$ in the typical
parameter region. For instance, its variation is smaller than $10^{-6}$ for $%
g$ varying from $6\kappa $ to $\kappa $. This result cannot be understood
naively from Eq. (5), from which one gets a phase of $e^{i0}$ only when $%
g\gg \kappa $. The reason for this discrepancy is that we have two addressed
cavity modes with symmetric effective detunings $\Delta =\pm g$, and their
joint effect makes the phase factor $e^{i0}$ very stable even if $g$ is
reduced to a value comparable with $\kappa $. The stability of the
conditional phase against variations of $g$ in the typical parameter region
is an important advantage of our scheme, as $g$ in current experiments
suffers significant random variation (roughly by a factor of $2$) due to
residual atomic motion \cite{mckeever03}.

\begin{figure}[tb]
\epsfig{file=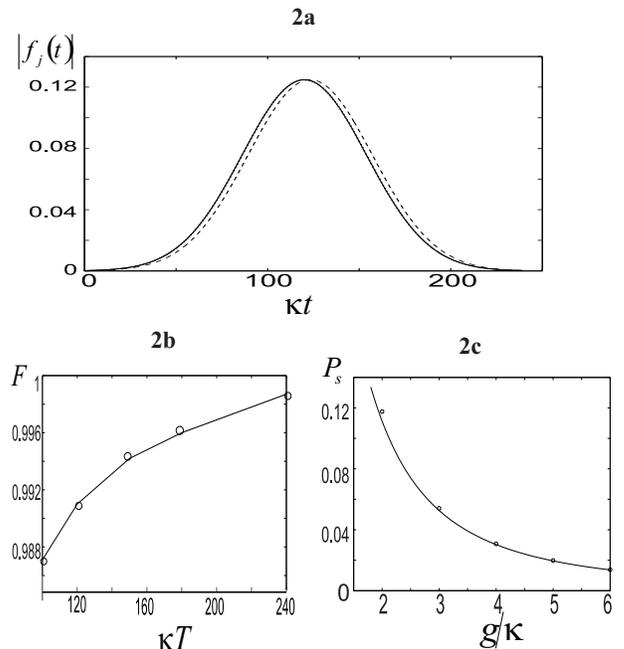,width=8cm}
\caption{(a) The shape functions $\left | f_{j}\left( t\right)
\right | $ for the input pulse (solid curve) and the reflected
pulse with the atom in the state $\left\vert 0\right\rangle $
(dashed curve) and $\left\vert 1\right\rangle $ (dotted curve),
respectively. The dotted and solid curves closely match and are
hardly distinguishable in the figure. (b) Fidelity $F$ due to
shape mismatch for the quantum CPF gate as a function of the input
pulse duration $T$ in units of $\protect\kappa ^{-1}$. The gate
fidelity quickly approaches $1$ for $\protect\kappa T\gg 1$. (c)
The probability $P_{s}$ of spontaneous emission loss versus the
normalized cavity coupling rate $g/\protect\kappa $, assuming
$\protect\gamma _{s}=\protect\kappa $ (circles). The solid curve
shows the fit by the empirical formula $P_{s}\approx 1/\left(
1+2g^{2}/\protect\kappa \protect\gamma_{s}\right) $. Other
parameters for (a), (b), $g=3\kappa$, $\protect\gamma_{s}
=\protect\kappa $, $\Delta=0$, and for (c), $T/5=24/\kappa$,
$\Delta=0$.} \label{fig2}
\end{figure}

The simulation also shows that the output pulse basically has the same shape
as the input pulse if the pulse duration $T\gg 1/\kappa $. Fig. 2a shows the
output pulse shapes $|f_{i}(t)|$ for the cases of the atomic states $\left|
1\right\rangle $ and $\left| 0\right\rangle $, respectively, and
demonstrates very good overlap with the input pulse shape shown in the same
figure. In more quantitative terms, we consider the fidelity $F$ of the CPF
gate $U_{aj}^{CPF}$ for the input atom-photon state $\left| \Phi
_{ai}\right\rangle \otimes \lbrack \left| \Psi _{pi}\right\rangle =(\left|
h\right\rangle +\left| v\right\rangle )/\sqrt{2}]$.{\it \ }Reductions in $F$
below unity are caused by shape mismatching between the input and the output
pulses and can be numerically calculated. Fig. 2b shows the gate fidelity $F$%
\ calculated in this way for different pulse durations $T$. For\ $%
T=240/\kappa $ (corresponding to a pulse width $T/5\sim 1\mu $s
for the parameters of Ref. \cite{mckeever03}), the gate fidelity
is about $99.9\%$. The shape of the output pulse is also very
insensitive to variation of the coupling rate $g$ in the typical
parameter region. For instance, the
relative shape change is smaller than $10^{-4}$ for $g$ varying from $%
6\kappa $ to $\kappa $.

The dominant noise in our CPF\ gate arises from photon loss due to atomic
spontaneous emission, leading to a vacuum-state output when the input is a
single-photon pulse. This noise yields a leakage error (also called an
erasure error) which means that the final state is outside of the qubit
Hilbert space $\{\left| h\right\rangle ,\left| v\right\rangle \}$ \cite
{preskill-notes}. Fig. 2c shows the probability $P_{s}$ of spontaneous
emission loss as a function of $g/\kappa $ for the input state $\left|
1\right\rangle \otimes \left| h\right\rangle $, assuming $\gamma _{s}=\kappa
$. The curve is well simulated by the empirical formula $P_{s}\approx
1/\left( 1+2g^{2}/\kappa \gamma _{s}\right) $. If the initial state of the
system is $\left| \Phi _{ai}\right\rangle \otimes \left| \Psi
_{pi}\right\rangle $, the average probability of the leakage error per $%
U_{aj}^{CPF}$ gate is given by $P_{e}=P_{s}/4$. In current experiments \cite
{mckeever03}, typically $\left( \kappa ,\gamma _{s}\right) /2\pi \approx
\left( 8,5.2\right) $ MHz, and $g/2\pi \approx 25$ MHz, which yields $%
P_{e}\approx 0.8\%$. With these parameters, a typical pulse width $%
T/5\approx 24/\kappa \approx 0.5$ $\mu $s. As the pulses $j$ and $k$ are
injected successively for the CPF\ gate $U_{jk}^{CPF}$, we need to introduce
a time delay of few $\mu $s between them. For demonstration-of-principle
experiments, this time delay can be routinely achieved through simple fiber
loops. To obtain longer time delay, atomic ensembles could be employed to
store photon pulses for several seconds \cite
{fleischhauer00,duan01,kuzmich03}.

Because the principal noise in our scheme is photon loss during gate
operations which is modeled as a leakage error, very efficient quantum error
correcting codes can be incorporated into this computation scheme to achieve
fault-tolerance \cite{preskill-notes}. For instance, a rough estimate in
Ref. \cite{knill00} shows that through concatenated coding, quantum
computation can tolerate leakage error at a percent level per gate, as
compared to the error threshold of about $10^{-5}$ for general quantum
errors \cite{preskill-notes}. The leakage error only affects the probability
to register a photon from each pulse and has no influence on the fidelity of
its polarization state if a photon is registered for each qubit (e.g.,
through QND or destructive measurements). So, leakage error induces small
inefficiency for each gate (at a level of a few percents), which is not
debilitating for experimental quantum computing up to dozens of CPF gates
even without quantum error correction.

In summary, we have shown that a cavity with a single-trapped atom,
conventionally used as a single-photon source, can be exploited to realize
scalable, fully-functional quantum computation. The proposed scheme is well
based on the state-of-the-art in cavity quantum electrodynamics, is robust
to various experimental sources of noise, and offers a promising approach to
the realization of large-scale fault-tolerant quantum computation.

This work was supported by the Michigan start-up fund, by the
FOCUS center, by the MURI Center for Quantum Networks (No.
DAAD19-00-1-0374), by the NSF (Nos. EIA-0086038 and PHY-0140355),
and by the Office of Naval Research (No. N00014-02-1-0828).


\end{document}